\documentclass[
  reprint,
  amsmath,amssymb,
  aps,
  pra,
  superscriptaddress,
  longbibliography
]{revtex4-2}
\usepackage[version=3]{mhchem} 
\usepackage{float}
\usepackage{comment}
\usepackage{esint}
\usepackage{tabularx}
\usepackage{soul}
\usepackage{amsmath}
\usepackage{amssymb}
\usepackage[normalem]{ulem}

\usepackage[utf8]{inputenc}
\usepackage[T2A,T1]{fontenc}
\usepackage[english]{babel}
\usepackage[
colorlinks=true,bookmarks=false,citecolor=blue,urlcolor=blue
]
{hyperref}

\usepackage{graphicx}
\usepackage{color}
\usepackage{soul}
\usepackage{xcolor}
\usepackage{comment}

\usepackage{xfrac}
\usepackage{gensymb}
\usepackage{multirow}
\usepackage[caption=false]{subfig}

\usepackage{dcolumn}
\usepackage{bm}

\begin{document}
\preprint{APS/123-QED}

\title{Engineering photonic crystal slab modes for strong exciton-photon coupling and polariton dispersion control}

\author{Natalia Salakhova}
\affiliation{Skolkovo Institute of Science and Technology, Moscow, Russia}

\author{Andrey Demenev}
\affiliation{Osipyan Institute of Solid State Physics RAS, Chernogolovka, Russia}

\author{Vladimir Kulakovskii}
\affiliation{Osipyan Institute of Solid State Physics RAS, Chernogolovka, Russia}

\author{Nikolay Gippius}
\affiliation{Skolkovo Institute of Science and Technology, Moscow, Russia}

\begin{abstract}
We numerically study exciton-photon coupling in a hybrid structure composed of a period-doubled Si$_3$N$_4$ photonic crystal slab and an hBN-encapsulated MoSe$_2$ monolayer. Period doubling folds quasi-guided modes into the light cone and produces spectrally separated photonic branches whose radiative character is controlled by the symmetry of the doubled unit cell, resulting in one bright and one bound state in the continuum (BIC)-like mode at normal incidence. We then use the hBN thickness as an independent tuning parameter for the vertical field profile. Increasing the hBN thickness redistributes the resonant field toward the MoSe$_2$ monolayer and enhances the Rabi splitting from about 3–4 meV to 8.5 meV, corresponding to an increase in the coupling strength from approximately 1.5–1.7 meV to 4.3–4.4 meV. Comparison between bright and BIC-like configurations shows that the quality factor primarily determines radiative visibility and linewidth, whereas the coupling strength is mainly governed by the field amplitude at the monolayer. We further show that the hBN layer controls the hybridization of folded TE-like modes at finite momentum, leading to an additional photonic gap whose size can be tuned by the hBN thickness and closes near $h_{\mathrm{hBN}}=113$ nm. In the hybrid structure, we observe finite momentum exciton-polariton states that, depending on the hBN thickness, form either through coupling to an isolated photonic branch or through coupling to two nearly degenerate photonic modes. These results identify period-doubled photonic crystal slabs as a flexible platform for controlling radiative coupling, exciton-photon overlap, and polariton dispersion in transition-metal dichalcogenides monolayers-based hybrid structures.
\end{abstract}
 
\maketitle

\section{Introduction}
Exciton - polaritons are hybrid light-matter excitations formed in the strong-coupling regime between optical modes and excitonic resonances. Their small effective mass inherited from the photonic component and their nonlinear response inherited from the excitonic component make them a versatile platform for nonlinear optics \cite{sturm2014all,suarez2021enhancement,demenev2019ultrafast,zasedatelev2021single,sanvitto2016road,kravtsov2020nonlinear}, polariton lasing \cite{azzini2011ultra,schneider2013electrically}, nonequilibrium condensation \cite{kasprzak2006bose,caputo2018topological,deng2010exciton,bloch2022non}, superfluidity \cite{amo2009superfluidity,lerario2017room,ma2026vortices}, and related collective phenomena.

In planar semiconductor microcavities, polariton dispersions are usually shaped by the cavity photon mode, while additional lateral potentials can be introduced through surface acoustic waves \cite{krizhanovskii2013effect,cerda2013exciton} or corrugated lateral confinement \cite{tanese2013polariton}. Such approaches have enabled polariton minibands, gap-confined states, and gap solitons.

Photonic crystal waveguides and slabs provide a different route to polariton band engineering \cite{PhysRevB.106.115424,nigro2023theory}. In these systems, periodicity is built into the optical structure itself. The grating folds guided modes into the light cone, allowing them to couple to external radiation while retaining their waveguide-like character \cite{tikhodeev2002}. Depending on the unit-cell symmetry and the in-plane wave vector, the folded modes can be radiatively bright, weakly leaky, or symmetry-protected at normal incidence. This makes photonic crystal slabs attractive for combining radiative accessibility with high-quality resonances, including bound states in the continuum (BIC), modes with potentially infinite lifetime. A distinctive feature of photonic crystal polaritons is the possibility to engineer the sign and magnitude of the effective mass, open gaps between folded branches, and create nearly flat or strongly nonparabolic dispersions. This is important not only for linear spectroscopy, but also for polariton condensation and nonlinear propagation in structures~\cite{kravtsov2020nonlinear,ardizzone2022polariton,dang2022realization,kim2021topological,riminucci2023polariton}. 

Recent experiments and theory have shown that BIC-like and gap-confined photonic modes can support polariton condensation and that the interplay between mode symmetry, effective mass, and radiative losses can strongly affect the condensation threshold~\cite{ardizzone2022polariton,riminucci2023polariton,nigro2023theory}. A supersolid polariton phase has also been predicted with close analogies to supersolidity in helium and ultracold atomic gases~\cite{nigro2025supersolidity}. 
Such a regime requires a specific finite-momentum dispersion, which is naturally accessible in photonic crystal structures through band folding and mode hybridization.

Many photonic-crystal polariton platforms studied so far are based on quantum wells embedded in patterned semiconductor waveguides~\cite{PhysRevB.106.115424,ardizzone2022polariton,riminucci2023polariton,nigro2023theory}. Quantum wells provide a mature material platform, but their relatively small exciton binding energies and the limited refractive-index contrast of lattice-matched semiconductor heterostructures can restrict the operating conditions and the achievable photonic band gaps.

Perovskites and organic semiconductors provide a route to room-temperature polariton platforms, owing to their large oscillator strengths~\cite{daskalakis2015spatial,dang2022realization,kim2021topological,zasedatelev2021single}. However, organic polariton systems often involve energetic disorder, broad vibronic absorption bands, inhomogeneous broadening, and many optically dark excitonic states, which can complicate spectral interpretation and affect coherent transport and condensation ~\cite{khazanov2023embrace,pandya2022tuning}.

Two-dimensional transition-metal dichalcogenide (TMD) monolayers and van der Waals aterials  offer an alternative excitonic medium \cite{dufferwiel2015exciton}. In the monolayer limit, materials such as MoSe$_2$, WS$_2$, and WSe$_2$ host direct excitonic transitions with large oscillator strengths and binding energies~\cite{laturia2018dielectric,fang2019control}. These properties make TMD monolayers and van der Waals semiconductors promising for hybrid photonic crystal platforms \cite{kravtsov2020nonlinear,tang2020ultrafast,weber2023intrinsic}. In such hybrid structures, the strength and observability of exciton-photon coupling are controlled by several parameters simultaneously. 

A high photonic quality factor reduces radiative decay, but it does not by itself ensure strong coupling to a two-dimensional excitonic sheet. The electric-field amplitude at the monolayer, the symmetry of the folded mode, and the radiative visibility of the resonance are also essential. The hBN layer, usually introduced as an encapsulation and spacer material in hybrid structures with TMD monolayers, can therefore play an optical role by changing the vertical waveguide geometry. This motivates a platform in which the photonic unit-cell symmetry and the vertical field profile can be tuned independently.

In this work, we numerically study a hybrid photonic-crystal platform consisting of a period-doubled Si$_3$N$_4$ slab coupled to an hBN-encapsulated MoSe$_2$ monolayer. The doubled in-plane periodicity folds quasi-guided modes into the light cone, increases the splitting between the folded photonic branches compared with the single-period geometry, and allows their radiative character to be controlled through the symmetry of the supercell. By comparing two modulation geometries, we show that the upper and lower folded branches can exchange their bright and BIC-like character. We then use the total hBN thickness as an independent parameter that modifies the vertical field profile and controls the exciton-photon coupling strength at the $\Gamma$ point. Finally, we show that the hBN thickness controls the hybridization of folded TE-like branches at finite in-plane momentum, enabling finite-$k_x$ polariton states formed either through an isolated photonic branch or through two nearly degenerate photonic modes. This provides a route to independently adjust radiative visibility, excitonic overlap, and polariton dispersion in TMD-photonic crystal structures.

\begin{figure}[h]
\centerline{\includegraphics[width=1\columnwidth,draft=false]{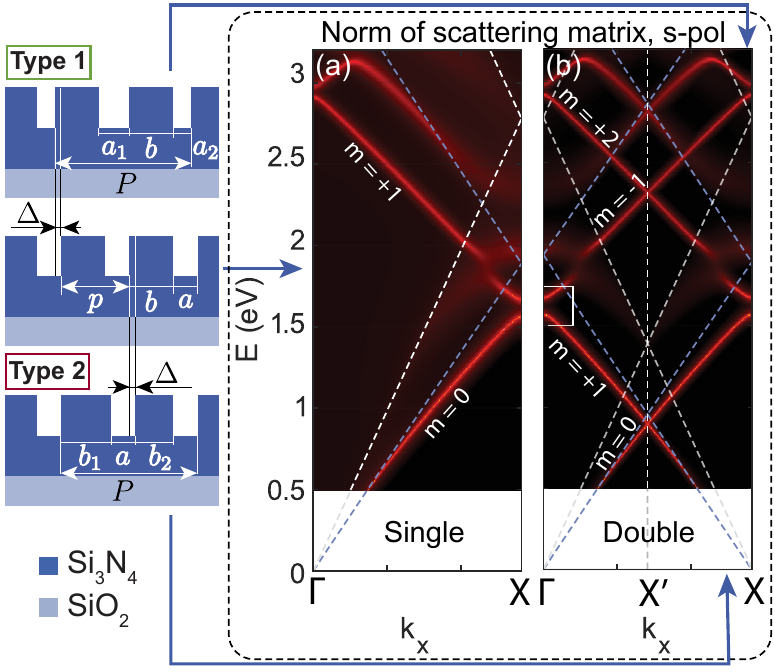}}
\caption{
Band folding in a one-dimensional photonic crystal slab.
Left: schematic of the initial grating with period $p$ and two period-doubled supercells with period $P=2p$: Type~1, where the groove width is modified, and Type~2, where the width of the dielectric stripes is modified.
Right: norm of the scattering matrix for s-polarized light as a function of the in-plane wave vector $k_x$ and photon energy for (a) the single-period and (b) the period-doubled structures.
The dashed lines indicate the light cones of air and substrate and the vertical dashed lines mark the Brillouin-zone boundaries.
}
\label{fig:pc_disp}
\end{figure}

\section{Model}
\label{sec:model}

We consider a highcontrast one-dimensional Si$_3$N$_4$ photonic crystal slab placed on a SiO$_2$ substrate. The structure is periodic along the $x$ direction, invariant along $y$, and layered along $z$. The total thickness of the Si$_3$N$_4$ slab is $h_{\mathrm{sl}}=266~\mathrm{nm}$. A grating is etched into the upper part of the slab with an etching depth $h_{\mathrm{gr}}=133~\mathrm{nm}$, leaving a homogeneous Si$_3$N$_4$ layer underneath as shown in Fig.~\ref{fig:pc_disp}. 

The initial grating period is $p=223~\mathrm{nm}$, with groove width $a=80~\mathrm{nm}$ and dielectric stripe width $b=143~\mathrm{nm}$. Period doubling is introduced by using a supercell with period $P=2p=446~\mathrm{nm}$, in which two neighboring grooves or stripes have different widths. In the \textbf{Type~1} structure, the widths of the two grooves in the supercell are changed as
$$
a_1=a+\Delta, \qquad a_2=a-\Delta,
$$
while the dielectric stripe widths remain equal to $b$, $\Delta$ is the modulation parameter. In the \textbf{Type~2} structure, the groove widths remain equal to $a$, whereas the dielectric stripe widths are changed as
$$
b_1=b+\Delta, \qquad b_2=b-\Delta.
$$
The corresponding geometries are shown in Fig.~\ref{fig:pc_disp} upper and lower relatively. In several calculations below, the grating period is varied to tune the photonic resonances with respect to the MoSe$_2$ exciton and all lateral dimensions are scaled so that the average groove filling factor and the modulation factor remain fixed:
$$
f=\frac{2a}{P}, \qquad f_{\Delta}=\frac{\Delta}{P}.
$$
The hybrid structure is formed by placing an hBN-encapsulated MoSe$_2$ monolayer on top of the photonic crystal slab. The lower hBN layer between the photonic crystal and the MoSe$_2$ monolayer is kept fixed at $h_{\mathrm{hBN1}}=3~\mathrm{nm}$, while the upper hBN layer thickness $h_{\mathrm{hBN2}}$ is varied. Throughout the Results section, we use the total hBN thickness as the parameter $h_{\mathrm{hBN}}=h_{\mathrm{hBN1}}+h_{\mathrm{hBN2}}$.

Si$_3$N$_4$ and SiO$_2$ are treated as isotropic dielectrics, with optical constants taken from Refs.~\cite{luke2015broadband,malitson1965interspecimen}. hBN is treated as a uniaxial anisotropic dielectric with the permittivity tensor components $\varepsilon_{x}^{\mathrm{hBN}}=\varepsilon_{y}^{\mathrm{hBN}}=2.89+0.001i$ and $\varepsilon_{z}^{\mathrm{hBN}}=4.97+0.001i$ at optical frequencies~\cite{laturia2018dielectric}.

The MoSe$_2$ monolayer is also modeled as an anisotropic layer whose out-of-plane component is taken as $\varepsilon_{z}^{\mathrm{MoSe_2}}=7.2$ whereas the in-plane components include the exciton resonance and are written as

\begin{equation}
\label{mose2}
\varepsilon_{x}^{\mathrm{MoSe_2}}(\omega)
=
\varepsilon_{y}^{\mathrm{MoSe_2}}(\omega)
=
\varepsilon_{\infty}^{\parallel}
+
\frac{\hbar\nu_{\rm X}}{E_{\rm X}-\hbar\omega-i\gamma_{\rm X}}.
\end{equation}

Here $\varepsilon_{\infty}^{\parallel}=16.8$ is the non-resonant in-plane background permittivity, $E_{\rm X}=1.642~\mathrm{eV}$ is the exciton resonance energy, $\hbar \nu_{\rm X}=33.6~\mathrm{meV}$ is the oscillator strength parameter, and $\gamma_{\rm X}=1~\mathrm{meV}$ is the excitonic decay rate used in the Lorentz oscillator model~\cite{fang2019control,laturia2018dielectric}. In the calculations where we analyze purely photonic modes, the resonant term is omitted and the MoSe$_2$ tensor is set to
$$
\hat{\varepsilon}^{\mathrm{MoSe_2}}=
\mathrm{diag}(16.8,16.8,7.2)+0.001i.
$$

We use the Fourier modal method, also known as rigorous coupled-wave analysis~\cite{tikhodeev2002,moharam1995}, which is well suited for layered periodic structures, to compute the scattering matrix $S(\omega,k_x)$ for given frequencies and in-plane wave vectors.
In a one-dimensional grating periodic along the $x$ direction, s- and p-polarized waves are decoupled as long as the $y$ component of the wave vector remains zero $k_y = 0$. In this study all configurations satisfy this condition and we can consider only s-polarized light separately. Transmission spectra and field distributions were calculated from the scattering matrix. The transmission maps correspond to the intensity $|t_s|^2$ of the zeroth transmitted diffraction order for s-polarized incidence and the dominant electric-field component is $\mathcal{E}_y$.


\section{Results}
\label{sec:results}

\begin{figure*}[t]
\centering
\includegraphics[width=0.75\linewidth]{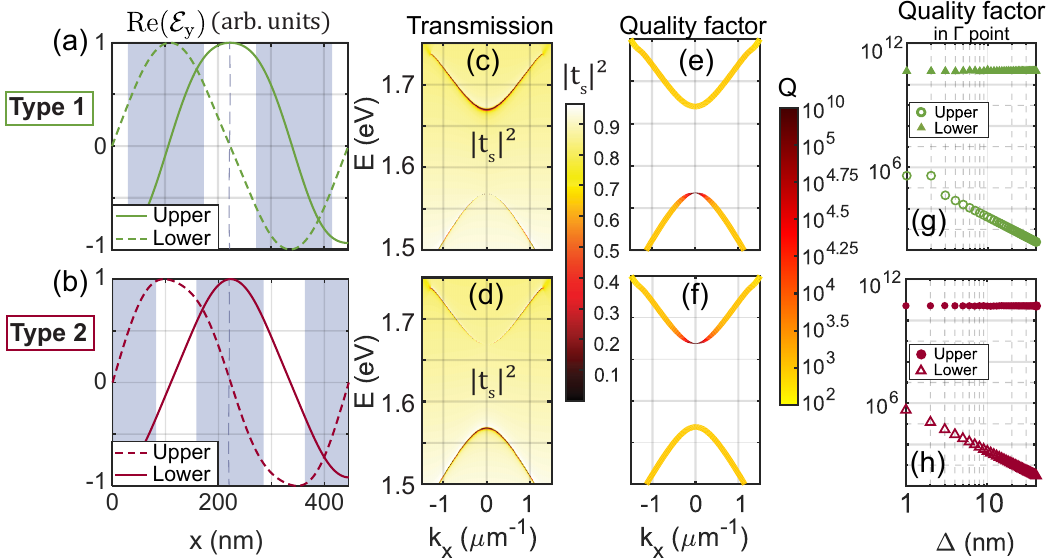}
\caption{Symmetry and radiative properties of quasi-guided modes in period-doubled photonic crystal slabs of Type 1 and Type 2.
(a,b) Real part of the dominant electric-field component $\mathcal{E}_y$ along the supercell at the $z$-coordinate corresponding to the middle of the grating.
(c,d) Transmission spectra as a function of the in-plane momentum.
(e,f) Quality factors of the corresponding modes. The line thickness and color reflect the line width and  quality factor.
(g,h) Quality factors of the upper and lower modes at the $\Gamma$ point as a function of the modulation parameter $\Delta$. }
\label{fig:pc_quality}
\end{figure*}

\subsection{Band folding in period-doubled photonic crystal slabs}

In contrast to conventional dielectric slab waveguides, photonic crystal slabs support quasi-guided modes that can couple to external radiation through the in-plane periodic modulation. This radiative coupling makes the modes visible in far-field transmission spectra while preserving their waveguide-like character. Such a combination is particularly useful for hybrid light-matter systems, where the optical mode must be both confined near the active material and experimentally observable.

The coupling between guided modes and external plane waves is enabled by the reciprocal lattice vectors of the grating, which relax the in-plane momentum conservation condition. As a result, guided-mode dispersions are folded into the first Brillouin zone. Figure~\ref{fig:pc_disp}(a) shows the norm of the scattering matrix for the initial one-dimensional photonic crystal slab  with period $p$ under s-polarized excitation. 

The bright branch corresponds to the lowest-energy TE mode that starts just under the light cone of the substrate (dashed light blue line). The folded guided-mode branches appear inside the light cone and become visible as resonant features in the scattering spectrum. The broad weakly visible resonance that starts under the light cone of the vacuum (white dashed line) and propagates to the boundary of the first Brillouin zone is the second TE mode. It is very broad due to strong leakage into the substrate and becomes much narrower after crossing the substrate light line near 2.5 eV.  

Introducing period doubling changes the supercell period to $P=2p$. This reduces the Brillouin zone by a factor of two and folds the guided modes with respect to the new zone boundary $\mathrm{X}'$, as shown in Fig.~\ref{fig:pc_disp}(b). The $m=+1$ mode in  panel (a) becomes a second diffraction order mode with $m=+2$. Period doubling allows to fold the modes from the initial Brillouin zone boundary X to the $\Gamma$ point preserving the spectral range of interest indicated by white rectangle in panel (b). 

Two types of period doubling are considered. In the Type~1 structure, the width of the groove is modified, whereas in the Type~2 structure the modulation is introduced through the widths of the dielectric stripes: the first has a mirror-symmetry plane in the air groove, whereas the second has it in the dielectric stripe (see gray dashed line in Fig.~\ref{fig:pc_quality}(a-b)). These two branches can be viewed as standing-wave combinations of counter-propagating guided waves coupled by the reciprocal lattice vector of the doubled grating, $G'=2\pi/P$. In the following, we refer to them as the upper and lower photonic or quasi-guided modes. They have a parabolic-like dispersion near the $\Gamma$ point, and their radiative characteristics are not fixed only by their spectral order, but depend on the symmetry of the supercell.

These symmetry-controlled radiative characteristics are analyzed in Fig.~\ref{fig:pc_quality}. Panels~(a) and~(b) present the real part of the dominant electric-field component, $\mathcal{E}_y$, along the unit cell for the upper and lower folded modes. For the Type~1 structure, the upper mode has even parity with respect to the symmetry plane of the supercell (dashed gray line), whereas the lower mode is odd. Since the electric field of a normally incident plane wave is uniform along $x$, only the even mode has a nonzero symmetry-allowed overlap with the external radiation channel. Consequently, the upper mode is bright and observable in the transmission spectrum, while the lower mode is dark at normal incidence, as shown in Fig.~\ref{fig:pc_quality}(c).

For the Type~2 structure, the position of the plane of mirror symmetry is shifted because the period doubling is introduced through the width of the dielectric stripes. This reverses the parity assignment of the folded modes. As a result, the lower branch becomes bright, whereas the upper branch is symmetry-protected and remains dark at the $\Gamma$ point, as shown in Figs.~\ref{fig:pc_quality}(b,d). Thus, by choosing the type of period doubling, we can select whether the upper or the lower photonic branch is radiatively accessible.

The dark modes at $k_x=0$ are symmetry-protected bound states in the continuum (BIC). In an ideal lossless periodic structure, their radiative quality factor is formally divergent; in the numerical calculations it is therefore limited by the precision of the resonance extraction. In our numerical calculations we add a very small imaginary part to the Si$_3$N$_4$ permittivity $\varepsilon"_{\rm Si_3N_4} = 10^{-10}$ to avoid numerical sensitivity of the resonance extraction to machine precision. When the in-plane wave vector becomes nonzero, the symmetry protection is lifted and these modes become leaky quasi-BICs with a rapidly decreasing quality factor, as shown in Fig.~\ref{fig:pc_quality}(e,f). In contrast, the bright modes have a finite radiative linewidth already at normal incidence. For the modulation depth $\Delta=20~\mathrm{nm}$, their quality factors are of the order of $10^3$, while the resonances remain clearly visible in transmission. 

The quality factor of the bright modes can be tuned over a wide range by changing the modulation depth. Figures~\ref{fig:pc_quality}(g,h) show that decreasing $\Delta$ suppresses the radiative coupling and leads to an inverse-square increase of $Q$, consistent with the quasi-BIC scaling expected for weak symmetry-breaking perturbations. For large modulation, $\Delta=40~\mathrm{nm}$, the quality factor is reduced to $Q\sim600$, whereas for $\Delta=20~\mathrm{nm}$ it is around $Q\sim2000$. Further reduction of the modulation increases the quality factor to $Q\sim10^4$ at $\Delta=10~\mathrm{nm}$ and to $Q\sim10^5$ at $\Delta=2~\mathrm{nm}$. The quality factor of the dark mode is nearly insensitive to the modulation parameter $\Delta$ as long as the symmetry of the unit cell is preserved.

An additional reduction of the quality factor is observed for the upper branches at finite in-plane momentum, near $k_x\simeq1.3~\mu\mathrm{m}^{-1}$. This feature is associated with the opening of an additional leakage channel when the mode crosses the folded substrate light cone.

Thus, period doubling provides two important advantages for the subsequent exciton-photon coupling study. First, the type of modulation controls whether the relevant folded branch is bright and directly observable in transmission or dark and symmetry-protected at normal incidence. Second, additional folding creates photonic branches with large spectral separation, which allows us to suppress unwanted multimode interactions and analyze exciton coupling to a selected photonic mode.

\subsection{Exciton-photon coupling at the $\Gamma$ point}

\begin{figure}[th]
\centering
\includegraphics[width=1\linewidth]{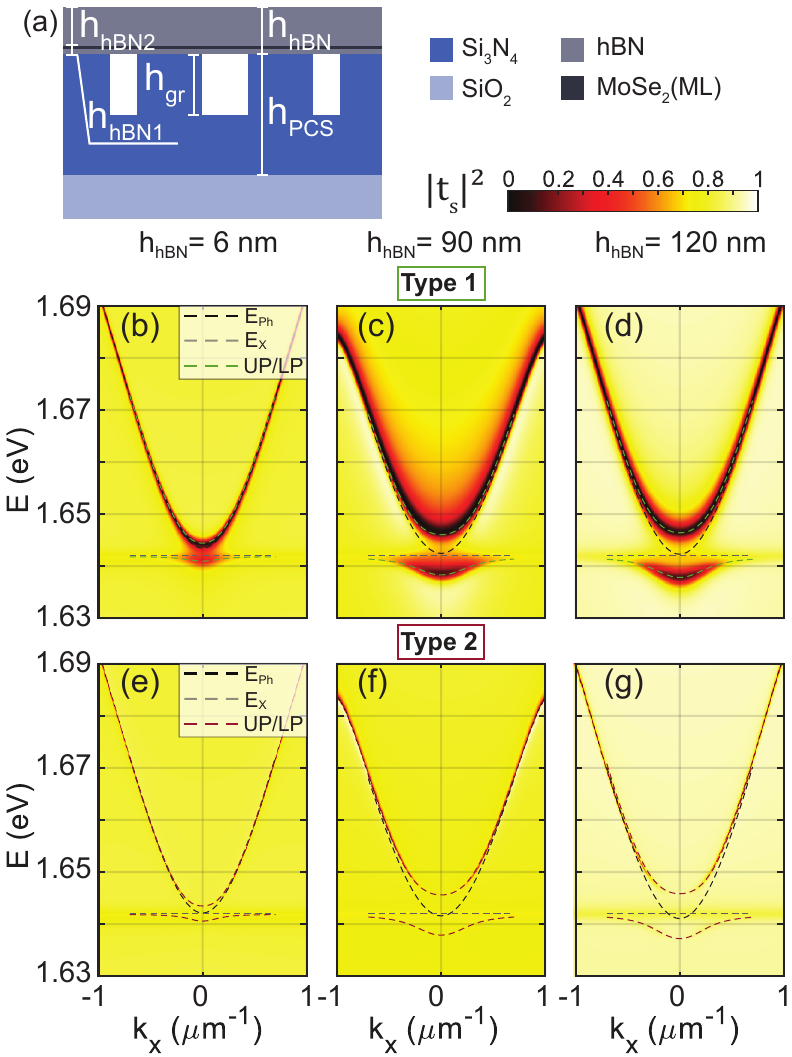}
\caption{Hybrid photonic crystal-hBN-MoSe$_2$ structure. (a) Schematic of the geometry: the MoSe$_2$ monolayer is encapsulated between a thin lower hBN layer and an upper hBN layer with variable thickness. (b-d) and (e-g) Transmission spectra of the Type~1 and Type~2 structures as a function of the in-plane momentum $k_x$. The black dashed lines indicate the energy of the initial photonic mode, the gray line indicates the exciton energy, and the green and dark-red lines indicate the upper and lower polariton branches. The optimal geometric parameters of the grating minimizing the photonic–exciton detuning: Type 1 $P_{\rm opt} = 454,~436.5,~ 422.5$~nm; Type 2 $P_{\rm opt} = 454,~435.5,~ 421.5$~nm; the filling factor $f = 0.37$ and modulation factor $f_{\Delta} = 0.046$.}
\label{fig:polariton}
\end{figure}

We next consider the hybrid structure formed by the period-doubled photonic crystal slab and a MoSe$_2$ monolayer encapsulated in hBN, as shown in Fig.~\ref{fig:polariton}(a). The lower hBN layer between the photonic crystal and the monolayer is kept fixed at $h_{\mathrm{hBN1}}=3~\mathrm{nm}$, while the upper hBN layer $h_{\mathrm{hBN2}}$ is varied. hBN encapsulation is commonly used to protect TMD monolayers and to provide a flat dielectric environment, but in the present structure we will show that it also acts as an optical tuning parameter that controls the vertical distribution of the resonant field. 

In this section, we consider the interaction between the exciton and the upper photonic mode with positive effective mass. In hybrid structure with a MoSe$_2$ , the interaction between the exciton and the photonic mode results in the formation of upper and lower polariton branches. Figures~\ref{fig:polariton}(b)-(d) show the transmission spectra for the Type~1 structure, where the relevant folded mode is bright and the polariton branches are directly visible. The Rabi splitting between upper and lower polariton branches increases with an increase of the hBN thickness. Figures~\ref{fig:polariton}(e)-(g) show the corresponding spectra for the Type~2 structure. In this case the relevant folded mode is dark at normal incidence and the polaritons inherit this property and are almost invisible in transmission because of the suppressed radiative leakage. Nevertheless, the extracted dispersions, shown by the dark red dashed lines in panels (e)-(g) are close to those obtained for the bright modes of Type~1 structure shown by the green lines in panels (b)-(d).

These dispersions were extracted as complex pole energies of the scattering matrix. We use the convention $\tilde{E}=E-i\gamma$, where $\gamma$ is the modal decay rate, or equivalently, the half-linewidth in energy units. For each hBN thickness and each period-doubling geometry, we first extract the complex energy of the uncoupled photonic mode, $\tilde{E}_{\mathrm{ph}} = E_{\mathrm{ph}} - i\gamma_{\mathrm{ph}}$. Its real part, $E_{\mathrm{ph}}$, is shown by the black dashed lines in Fig.~\ref{fig:polariton}. We then calculate the complex energies of the two polaritonic modes in the hybrid photonic crystal-MoSe$_2$ structure, $\tilde{E}_{\mathrm{UP}}$ and $\tilde{E}_{\mathrm{LP}}$.
Their real parts, $E_{\mathrm{UP}}$ and $E_{\mathrm{LP}}$, are shown by the colored lines.

Within the zero-detuning ($E_{\rm ph} = E_{\rm X} = E_0$) non-Hermitian two-oscillator model 
\begin{equation}
   \hat{H} =  \begin{pmatrix}
        E_{\rm ph} - i \gamma_{\rm ph} & g\\
        g^\dagger & E_{\rm X} - i \gamma_{\rm X}
    \end{pmatrix}
\end{equation}
yields two eigenenergies
\begin{equation}
    \tilde{E}_{\rm UP/LP} = E_0-i\frac{\gamma_{\rm ph}+\gamma_{\rm X}}{2}{}\pm\sqrt{g^2-\big(\frac{\gamma_{\rm ph}-\gamma_{\rm X}}{2}\big)^2}
\end{equation}
that follows the Rabi splitting:
\begin{equation}
\label{rabi}
    \Omega_R =  E_{\rm UP}-E_{\rm LP} = 2\sqrt{g^2-\big(\frac{\gamma_{\rm ph}-\gamma_{\rm X}}{2}\big)^2}
\end{equation}
The maximum achievable value of splitting between polariton branches is $\Omega_R^{\max} = 2g$ and from Eq. \ref{rabi}

\begin{equation}
    g=\frac{1}{2}\sqrt{\Omega_R^2+\left(\gamma_{\mathrm{ph}}-\gamma_{\rm X}\right)^2}.    
\end{equation}

The Rabi splitting values are extracted from the calculated complex pole energies, with the corresponding real parts shown in Fig.~\ref{fig:polariton}. The photonic decay rate is obtained from the imaginary part of the uncoupled photonic pole, $\gamma_{\mathrm{ph}}=-\mathrm{Im}\,\tilde{E}_{\mathrm{ph}}$, whereas the excitonic decay rate is taken as $\gamma_{\mathrm{X}}=1~\mathrm{meV}$, consistent with the parameter used in the Lorentz oscillator model for the MoSe$_2$ monolayer (Eq.~\eqref{mose2}) and with the low-temperature linewidth reported for hBN-encapsulated MoSe$_2$ monolayers~\cite{fang2019control}. We then use these quantities to calculate the coupling parameter $g$ for each relevant geometry. The results for each type of structure are summarized in Table~\ref{tab:gamma_coupling}. 

\begin{table}[h]
\renewcommand{\arraystretch}{1.3}
\caption{Exciton-photon coupling strength at the $\Gamma$ point for Type 1 and Type 2 structures.
The Rabi splitting, photonic and excitonic linewidth $\gamma_{\rm ph}$ and $\gamma_{\rm X}$ and coupling energy $g$ are presented in meV.}
\label{tab:gamma_coupling}
\begin{center}
\begin{tabular*}{0.95\linewidth}{@{\extracolsep{\fill}}|c|c|c|c|c|c|}
\hline
$h_{\rm hBN}$ 
& Type 
& $\Omega_R$ 
& $\gamma_{\rm ph}$ 
& $\gamma_{\rm X}$ 
& $g$ 
\\
\hline
\hline
\multirow{2}{*}{6}
& 1 & 3.435 & 0.826   & 1.000 & 1.720 \\
& 2 & 2.897 & 0.00006 & 1.000 & 1.532 \\

\hline

\multirow{2}{*}{90}
& 1 & 7.779 & 2.201 & 1.000 & 3.936 \\
& 2 & 7.824 & 0.005 & 1.000 & 3.944 \\

\hline

\multirow{2}{*}{120}
& 1 & 8.668 & 2.297 & 1.000 & 4.382 \\
& 2 & 8.616 & 0.010 & 1.000 & 4.336 \\

\hline
\end{tabular*}
\end{center}
\end{table}

For both geometries, $g$ increases from approximately $1.5$-$1.7~\mathrm{meV}$ at $h_{\mathrm{hBN}}=6~\mathrm{nm}$ to $4.3$-$4.4~\mathrm{meV}$ at $h_{\mathrm{hBN}}=120~\mathrm{nm}$. At $h_{\mathrm{hBN}}=90$ and 120~nm, correction associated with the difference between $\gamma_{\mathrm{ph}}$ and $\gamma_{\rm X}$ is small compare to the polariton splitting. In this case the $g\simeq\Omega_R/2$ and the Rabi splitting is close to achievable maximum. For the Type~2 structure at $h_{\mathrm{hBN}}=6~\mathrm{nm}$, the photonic mode remains nearly BIC-like and $\gamma_{\mathrm{ph}}\ll\gamma_{\mathrm{X}}$. The finite excitonic decay rate therefore gives a noticeable correction to the simple estimate $g\simeq\Omega_R/2$, so that the observable polariton splitting is smaller than the maximum achievable splitting $2g$. 

All considered cases satisfy the resolvability condition
$$
\Omega_R>\gamma_{\mathrm{ph}}+\gamma_{\rm X},
$$
which supports their assignment to the strong-coupling regime within the present linear model. The values in Table~\ref{tab:gamma_coupling} are given for the zero-detuning assumption. The same analysis including the detuning between the uncoupled photonic mode and the exciton is provided in the Supplementary Information and leads to the same conclusions.

The exciton-photon coupling increases with the hBN thickness even though the excitonic decay rate is kept fixed and the photonic decay rate even increase. This indicates that the enhancement is not caused by a narrowing of the photonic resonance. To identify its origin, we suppress the resonant part of the MoSe$_2$ permittivity and use the nonresonant dielectric tensor $\hat{\varepsilon}_{\mathrm{MoSe}_2}=\mathrm{diag}(16.8,16.8,7.2)$ to study how hBN changes the photonic-mode properties. 

\begin{figure}[th]
\centering
\includegraphics[width=1\linewidth]{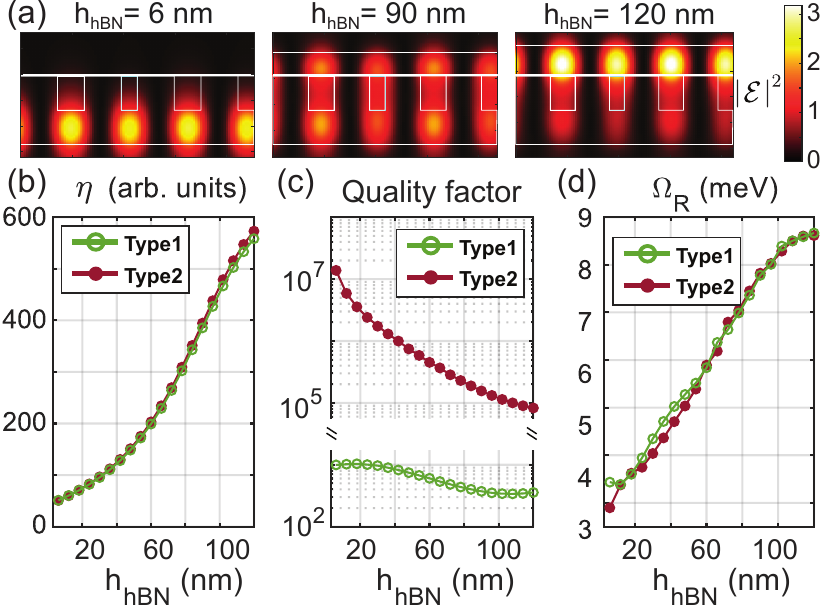}
\caption{(a) Eigenmode electric-field intensity $|\mathcal{E}(x,z)|^2$ distribution in $xz$ cross sections at $k_x,k_y = 0$ for the non-resonant MoSe$_2$ response and for three hBN thicknesses: $h_{\mathrm{hBN}}=6$, 90, and 120 nm. The field distributions are shown for the Type~1 structure. (b) Field-intensity parameter $\eta$ (Eq.~\ref{mose2_field}) at the MoSe$_2$ monolayer. (c) Quality factor of the photonic mode for Type~1 and Type~2 structures. (d) Rabi splitting $\Omega_R$ calculated for hybrid structure with resonant MoSe$_2$ response. The structure period for each field map in (a) and for each data point in (b-d) was chosen to shift the initial $E_{\rm ph}$ at the $\Gamma$ point as close as possible to $E_{\rm X}$. The filling factor $f = 0.37$ and modulation factor $f_{\Delta} = 0.046$.}
\label{fig:ex_hBN}
\end{figure}

Figures~\ref{fig:ex_hBN}(a) shows the corresponding eigenmode field-intensity distributions for the Type~1 structure at three representative hBN thicknesses. We analyze the eigenmode electric field rather than the scattered field, because the coupling strength describes how the excitonic dipole couples to the eigenmodes of the structure, independent of a particular external excitation channel. For $h_{\mathrm{hBN}}=6~\mathrm{nm}$, the mode is mainly localized inside the homogeneous part of the Si$_3$N$_4$ slab, with only a weak field tail reaching the hBN/MoSe$_2$ region. Increasing the hBN thickness redistributes the mode in the vertical direction. At $h_{\mathrm{hBN}}=90~\mathrm{nm}$, the field develops two pronounced maxima, one in the photonic crystal slab and the other in the hBN layer. At $h_{\mathrm{hBN}}=120~\mathrm{nm}$, the upper maximum becomes stronger, which increases the field intensity at the MoSe$_2$ monolayer position. Thus, the exciton–photon coupling increases when the electric-field maximum shifts closer to the excitonic monolayer. 

To quantify this trend, we compare how field intensity, quality factor and Rabi splitting change with the thickness of the hBN layer.
We calculate the field-intensity parameter
\begin{equation}
\label{mose2_field}
   \eta=\int\limits_{0}^{P}\left|\mathcal{E}(x,z=z_{\mathrm{ML}})\right|^2 dx 
\end{equation}
that shows the average intensity of the electric field of the photonic mode in the region where the monolayer is located. Fig.~\ref{fig:ex_hBN}(b) shows nearly the same dependence on $h_{\mathrm{hBN}}$ for the two Types of geometries because the eigenmode distribution is mainly determined by the structure geometry rather than by the coupling to the external radiation continuum. 

Figure~\ref{fig:ex_hBN}(c) shows the quality factors of the Type~1 and Type~2 modes. The quality factor of the dark mode monotonically decreases because the modal overlap with the weakly absorbing hBN and MoSe$_2$ layers increases whereas its radiative losses remain negligible at normal incidence. For the bright mode, the dependence is nonmonotonic because of the interplay between radiative and nonradiative losses. The radiative contribution depends on the overlap with the grating region, which couples the mode to the radiation continuum and decreases when the mode is predominantly localized in Si$_3$N$_4$ or hBN homogeneous layers. The nonradiative contribution is associated with absorption and increases as the modal overlap with the absorbing layers grows. 

The Rabi splitting shown in Figure~\ref{fig:ex_hBN}(d) follows the same trend as $\eta$, increasing from about $3$-$4~\mathrm{meV}$ for thin hBN to about $8.5~\mathrm{meV}$ for $h_{\mathrm{hBN}}\approx120~\mathrm{nm}$. As discussed above, the coupling strength can be estimated as $g\simeq\Omega_R/2$, so this correlation indicates that the coupling strength is governed mainly by the field amplitude at the excitonic monolayer, whereas the quality factor controls the radiative decay rate and spectral visibility and has only a minor influence on the coupling strength.

For this reason, in the following we focus on the bright Type~1 structure. It provides nearly the same exciton-photon coupling strength as the dark Type~2 configuration, while the polariton branches remain directly observable in the transmission spectra.

\subsection{hBN-controlled photonic gap at finite in-plane momentum}
\begin{figure}[h]
\centering
\includegraphics[width=0.89\linewidth]{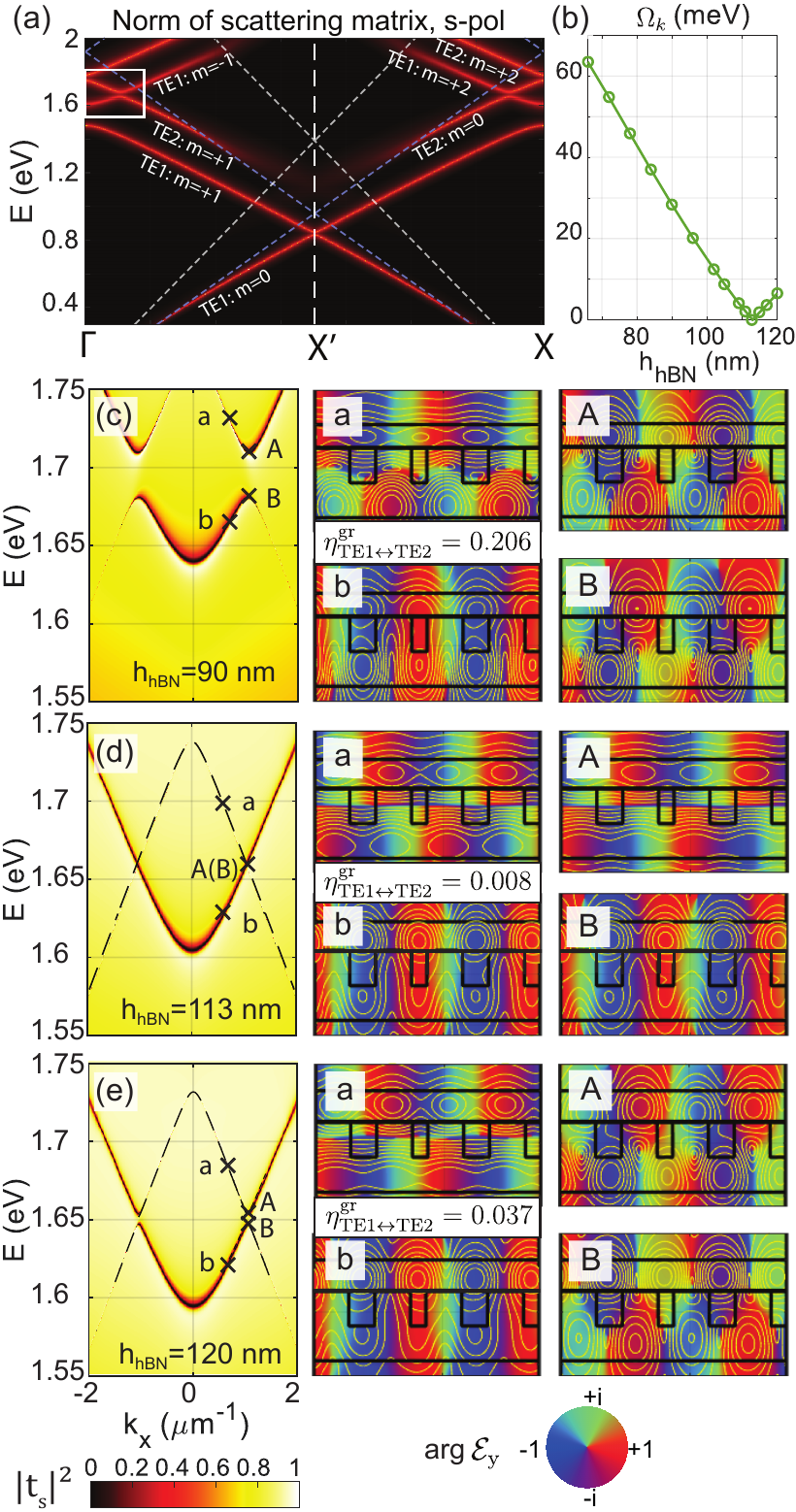}
\caption{Hybridization of quasi-guided modes. (a) Norm of the scattering matrix for the hybrid photonic crystal-hBN-MoSe$_2$ structure without excitonic resonance, showing the folded TE$_1$-like and TE$_2$-like guided modes.
(b) Finite-momentum photonic splitting $\Delta\Omega_k$ as a function of total hBN thickness.
(c-e) Transmission spectra as a function of $k_x$ and $xz$ maps of the eigenmode electric-field component $\mathcal{E}_y$ at four points \textbf{a}, \textbf{b}, \textbf{A}, \textbf{B} for $h_{\mathrm{hBN}}=90$, 113, and 120 nm. The color represents $\arg \mathcal{E}_y$, and the contour lines indicate the field amplitude levels. Parameters of the structure: $P=437$~nm, $a = 80$~nm, $\Delta = 20$~nm ($f = 0.37$ and $f_{\Delta} = 0.046$).}
\label{fig:gap_evolution}
\end{figure}

The hBN layer not only controls the field distribution but also modifies the dispersion of the quasi-guided modes, resulting in the opening of an additional photonic gap. We first analyze this effect in the structure where the excitonic contribution to the MoSe$_2$ permittivity is switched off. In the spectral range of interest, the hybrid photonic-crystal--hBN-MoSe$_2$ structure supports two TE-like guided modes, denoted below as TE1-like and TE2-like modes. Period doubling folds these modes into the reduced Brillouin zone, where the corresponding branches cross at finite $k_x$, as shown in Fig.~\ref{fig:gap_evolution}(a) (crossing region is highlighted by the white rectangle). The photonic gap originates from the hybridization of these folded TE$_1$-like and TE$_2$-like modes mediated by the periodic modulation of the photonic crystal slab
and mode splitting $\Omega_k$ depends on the thickness of the hBN layer.

To quantify the evolution of the gap, we extract the complex pole energies of the two interacting photonic modes and define the photonic splitting $\Omega_k$ as the difference between the real parts of the corresponding poles at the avoided crossing. This splitting is a purely photonic feature that should be distinguished from the exciton-photon Rabi splitting discussed above. As shown in Fig.~\ref{fig:gap_evolution}(b), $\Omega_k$ exceeds $60~\mathrm{meV}$ for $h_{\mathrm{hBN}}$ smaller than 70~nm and decreases as the hBN thickness increases. 

To clarify the physical origin of the gap closing, we compare three structures with hBN thicknesses $h_{\mathrm{hBN}}=90$, 113, and 120 nm. The transmission spectra in Figure~\ref{fig:gap_evolution}(c-e) show that for $h_{\mathrm{hBN}}=90~\mathrm{nm}$ the photonic gap is about $20~\mathrm{meV}$, at $h_{\mathrm{hBN}}=113~\mathrm{nm}$ the gap closes, and for $h_{\mathrm{hBN}}=120~\mathrm{nm}$ it reopens and reaches about $8~\mathrm{meV}$. This nonmonotonic behavior indicates that the hBN thickness changes not only the resonant energy of the modes but also the effective coupling between the TE$_1$- and TE$_2$-like modes.

This can be seen from the field distributions of modes for different hBN thicknesses. The field maps in Fig.~\ref{fig:gap_evolution}(c-e) show the phase and amplitude of the $y$-component of the electric field $\mathcal{E}_y$, in the $xz$ cross section at four selected points. The points \textbf{a} and \textbf{b} are chosen far from the crossing. The energy difference between modes in these points is around 70 meV. This energy separation allows us to treat the modes as nearly noninteracting and panels \textbf{a} and \textbf{b} represent nearly pure TE$_1$ and TE$_2$ modes. For the upper branch at point \textbf{a}, the field has two separated maxima, one in the homogeneous part of the Si$_3$N$_4$ slab and the other in the hBN layer. The field amplitude passes through a node near the grating region, accompanied by a phase jump at $\pi$, which is consistent with its assignment as a TE$_2$-like mode. For the lower branch at point \textbf{b}, the field does not exhibit such nodes, which is consistent with its TE$_1$-like character. These features are common for all three considered hBN thicknesses.

The points $A$ and $B$ correspond to the modes near the crossing or anti-crossing region. The behavior near the crossing region depends strongly on $h_{\mathrm{hBN}}$. For $h_{\mathrm{hBN}}=90$ and $120~\mathrm{nm}$, the fields at points \textbf{A} and \textbf{B} are mixed combinations of the modes observed at points \textbf{a} and \textbf{b}, which is consistent with the opening of a gap. In contrast, for $h_{\mathrm{hBN}}=113~\mathrm{nm}$, the field distributions at \textbf{A} and \textbf{B} remain close to those of the nearly pure modes at points \textbf{a} and \textbf{b}. This indicates that mode hybridization is strongly suppressed at this thickness.

As a qualitative measure of the coupling between the two modes, we also calculate the overlap parameter
$$
\eta^{\mathrm{gr}}_{\mathrm{TE1,TE2}} =
\frac{
\left|
\int_{V_{\mathrm{gr}}}
\mathbf{\mathcal{E}}_{a}^{*}\cdot\hat{\varepsilon}(x,z)\mathbf{\mathcal{E}}_{b}\,dV
\right|
}{
\sqrt{
\int_{V_{\mathrm{uc}}}
\mathbf{\mathcal{E}}_{a}^{*}\cdot \mathbf{\mathcal{E}}_{a}\,dV
}
\sqrt{
\int_{V_{\mathrm{uc}}}
\mathbf{\mathcal{E}}_{b}^{*}\cdot\mathbf{\mathcal{E}}_{b}\,dV
}
}.
$$
Here $V_{\mathrm{gr}}$ denotes the grating region of the unit cell, and $V_{\mathrm{uc}}$ denotes the full unit-cell volume. This parameter is not used as an absolute coupling coefficient, but it provides a useful relative measure of the overlap. For $h_{\mathrm{hBN}}=90$, 113, and 120 nm, we obtain $\eta^{\mathrm{gr}}_{\mathrm{TE1,TE2}}\approx0.206$, $0.008$, and $0.037$, respectively. Thus, at $h_{\mathrm{hBN}}=113~\mathrm{nm}$, the overlap parameter is reduced by approximately a factor of 26 compared with the $90~\mathrm{nm}$ case and by approximately a factor of 5 compared with the $120~\mathrm{nm}$ case.

These results show that the hBN thickness provides a direct way to tune the photonic splitting at finite in-plane momentum. The closing of the gap near $h_{\mathrm{hBN}}\approx113~\mathrm{nm}$ is consistent with a suppression of the grating-mediated coupling between the folded TE$_1$-like and TE$_2$-like modes.

\subsection{Finite-momentum exciton-polariton states}

\begin{figure}[th]
\centering
\includegraphics[width=0.8\columnwidth]{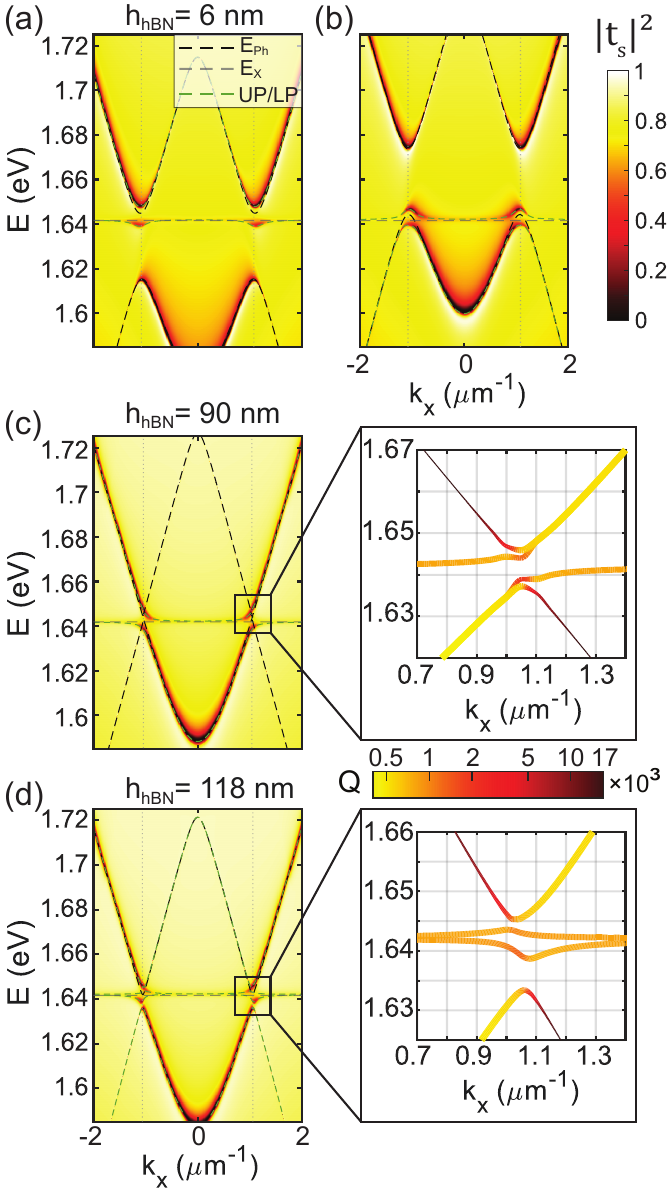}
\caption{Finite-momentum polaritons in the hybrid photonic crystal-hBN-MoSe$_2$ structure. Transmission spectra as a function of $k_x$ for different geometry parameters: (a) $P = 459$~nm and $h_{\rm hBN} = 90$~nm; (b) $P = 449$~nm and $h_{\rm hBN} = 90$~nm; (c) $P = 443$~nm and $h_{\rm hBN} = 113$~nm and (d) $P = 441$~nm and $h_{\rm hBN} = 118$~nm. The filling factor $f = 0.37$ and modulation factor $f_{\Delta} = 0.046$ are constant. the black dashed lines show the real parts of the initial photonic-mode energies $E_{\mathrm{ph}}$, the gray dashed lines indicate the exciton energy $E_{\rm X}$, and the green lines show the extracted upper and lower polariton branches UP and LP. The insets on the right-hand side of the panels (c) and (d) show the dispersion of the polariton modes where the line thickness and color indicate the linewidth and quality factor of the resonances.}
\label{fig:finite_k_pol}
\end{figure}

After establishing the hBN-controlled photonic gap at finite in-plane momentum, we include the resonant excitonic response of the MoSe$_2$ monolayer and tune the structure geometry so that the exciton intersects the new photonic branches near this gap. Through hBN-thickness control of the photonic gap size we can realize exciton–photon coupling either to a single photonic mode Fig.~\ref{fig:finite_k_pol}(a-b) or to two photonic modes simultaneously Fig.~\ref{fig:finite_k_pol}(c-d).

Figure~\ref{fig:finite_k_pol}(a,b) shows transmission spectra for two configurations with $h_{\rm hBN} = 90$~nm when the photonic gap is sufficiently large to treat the exciton as coupled mainly to a single photonic mode. The grating period is chosen so that the MoSe$_2$ exciton is brought close to either the upper or the lower photonic branch. The corresponding uncoupled photonic branch $E_{\mathrm{ph}}$, the exciton energy $E_{\rm X}$, and the extracted polariton branches are shown by dashed lines.

For cases shown in Figs.~\ref{fig:finite_k_pol} (c) and (d) for $h_{\rm hBN} = 113$~nm and $h_{\rm hBN} = 118$~nm where the initial photonic branches cross or mode splitting is small (around 4 meV), the exciton interacts simultaneously with two photonic modes. The insets on the right-hand side of panels (c) and (d) show the eigenmode dispersion, where the line thickness and color indicate the linewidth and quality factor of the mode. The initial photonic modes have different symmetries that results in different radiative losses. So the TE$_1$-like mode is leaky, whereas the TE$_2$-like mode is nearly dark. As a result, both the hybrid photonic branches and the polariton modes are asymmetric with respect to the crossing momentum $k_x$ in both their dispersion slopes and quality factors of the polariton modes.   

\section{Discussion}
\label{sec:disscusion}

The comparison between the bright and BIC-like configurations clarifies the different roles of radiative losses and field overlap in the strong-coupling regime. In the linear response considered here, a high photonic quality factor does not by itself increase the exciton-photon coupling strength. The coupling is mainly determined by the in-plane electric-field amplitude sampled by the monolayer. The quality factor instead controls the photon lifetime and the visibility of the resonance in far-field spectra. BIC-like character should therefore not be interpreted as a direct mechanism for increasing $g$. If the goal is to maximize the observable polariton splitting in transmission spectra, a nearly BIC-like mode can also be disadvantageous: its weak coupling to the external radiation continuum complicates direct observation, while a large imbalance between the excitonic and photonic decay rates reduces the observable splitting relative to the maximum value $2g$. Increasing $g$ requires enhanced spatial overlap between the optical mode and the excitonic monolayer, so that the excitonic dipole moment couples more efficiently to the electric field of the photonic eigenmode.

The bright and BIC-like regimes are therefore useful for different experimental and physical tasks. Bright folded modes are directly accessible in angle-resolved transmission or reflection and provide a convenient route for extracting polariton dispersions, linewidths, and Rabi splittings. This makes them preferable for the analysis of the linear spectra and motivates our focus on the Type 1 configuration in this work. BIC-like modes, in contrast, are weakly visible at normal incidence because their radiative coupling is forbidden by symmetry. They can still be accessed at finite in-plane momentum, by weak symmetry breaking, finite numerical aperture, or disorder-induced leakage. Their main advantage is the reduced radiative decay rate, which can be important for nonlinear polariton physics, including condensation and lasing thresholds. Thus, the bright configuration is more suitable for far-field spectroscopy, whereas the BIC-like configuration is preferable when long photon lifetime is required.

 The finite-momentum polariton states considered in Fig.~\ref{fig:finite_k_pol} differ from the usual $\Gamma$-point polaritons in two aspects. First, the exciton–photon crossings occur at finite and opposite in-plane wave vectors, $+k_x$ and $-k_x$, as required by the reciprocity of the passive photonic structure. Second, the number of photonic branches participating in the interaction depends on the hBN-controlled photonic gap. When the gap is large, the exciton predominantly couples to an isolated photonic branch. When the photonic gap is closed or small, two photonic modes are nearly degenerate near the exciton energy, and the exciton interacts with both of them. This produces a more complex polariton dispersion with extremely asymmetric linewidth variation inherited from the different radiative characters of the underlying photonic modes.

 The present calculations are restricted to the linear optical response and therefore do not address polariton condensation, nonlinear mode competition, or spontaneous formation of coherent finite-momentum states. Nevertheless, the finite-$k_x$ branches provide pairs of opposite-momentum states. Excitation of one momentum state at sufficiently high density could, in principle, seed scattering into the counterpropagating state through polariton-polariton interactions or four-wave-mixing processes. A coherent occupation of both opposite-momentum states is one of the ingredients discussed in the literature for supersolid polariton phases in photonic crystal waveguides \cite{nigro2025supersolidity}. Our results do not demonstrate such a phase, but they show that hBN-controlled hybrid photonic crystal slabs can provide a tunable finite-momentum dispersion required for this type of nonlinear physics.

Regarding experimental observation, several points should be noted. The hBN thicknesses considered here are compatible with deterministic transfer of hBN-encapsulated TMD monolayers, but the finite-$k_x$ gap closing near $h_{\mathrm{hBN}}\approx 113$ nm makes this regime sensitive to thickness variations. A fabrication-tolerance analysis would therefore be required for a direct experimental implementation. The coupling strength could also be increased by incorporating several TMD monolayers separated by thin hBN spacers and positioned within the high-field region of the mode. In this case, the collective oscillator strength of the excitonic subsystem would increase while the total thickness of the active region remains deeply subwavelength. This strategy would require spectral alignment of the excitons in different monolayers, together with control over inhomogeneous broadening, interlayer coupling, and additional absorption losses. Another limitation is the fixed excitonic linewidth used in the Lorentz oscillator model. The value $\gamma_X=1$ meV corresponds to a low-temperature, high-quality hBN-encapsulated MoSe$_2$ monolayer and should not be interpreted as a room-temperature parameter. Disorder, finite lateral size, exciton inhomogeneous broadening, and possible absorption losses in real materials are also not included explicitly. These effects may modify the linewidths and visibility of the polariton branches, but they are not expected to change the main design principle: the unit-cell symmetry controls the radiative character, whereas the hBN thickness controls both the excitonic overlap and finite-momentum photonic modes hybridization.

\section{Conclusion}
\label{sec:conc}

We have shown that period doubling in a one-dimensional Si$_3$N$_4$ photonic crystal slab provides symmetry control over folded quasi-guided modes. The two modulation geometries considered here reverse the symmetry assignment of the upper and lower folded branches, so that the radiatively bright and BIC-like modes exchange their spectral order. This gives a direct way to select whether a given photonic branch is accessible in far-field spectra or protected from radiative leakage at normal incidence.

For the hBN-encapsulated MoSe$_2$ monolayer, the total hBN thickness acts as an optical tuning parameter. Increasing $h{\mathrm{hBN}}$ shifts the resonant field toward the monolayer and increases the Rabi splitting from about $3–4$ meV to approximately $8.5$ meV. The extracted coupling strength increases from $g\approx 1.5–1.7$ meV to $g\approx 4.3–4.4$ meV. The comparison between bright and BIC-like configurations shows that the quality factor mainly controls the linewidth and far-field visibility, whereas the coupling strength is set primarily by the electric-field amplitude at the monolayer.

The hBN thickness also modifies the interaction between folded TE-like modes at finite in-plane momentum. The corresponding photonic gap decreases with increasing $h_{\mathrm{hBN}}$, closes near $h_{\mathrm{hBN}}=113$ nm, and reopens for larger thicknesses. This determines whether the MoSe$_2$ exciton couples to an isolated photonic branch when the gap is sufficiently large or to two nearly degenerate photonic modes when the gap is small or closed, resulting in finite-momentum polaritons with different dispersions.

These results show that period-doubled photonic crystal slabs provide a compact platform for independently adjusting radiative visibility, excitonic overlap, and polariton dispersion in TMD-based hybrid structures. Such control can be useful for finite-momentum polariton spectroscopy and for future studies of nonlinear polariton states in photonic-crystal geometries.

\section{Acknowledgement}

This work was supported by the Russian Science Foundation (Grant No. 24-12-00411).

 \nocite{*}
 \bibliographystyle{jpier}
 \bibliography{Ref}

\newpage

 \title{Supplementary Information for "Engineering photonic crystal slab modes for strong exciton-photon coupling and polariton dispersion control"}

\maketitle

Non-Hermitian two-oscillator model:

\begin{equation}
   \hat{H} =  \begin{pmatrix}
        E_{\rm ph} - i \gamma_{\rm ph} & g\\
        g^\dagger & E_{\rm X} - i \gamma_{\rm X}
    \end{pmatrix}
\end{equation}

In zero-detuning limit:
\begin{equation}
    \Omega_R = 2\sqrt{g_0^2-\big(\frac{\gamma_{\rm ph}-\gamma_{\rm X}}{2}\big)^2}
\end{equation}
and 
\begin{equation}
    g_0=\frac{1}{2}\sqrt{\Omega_R^2+\left(\gamma_{\mathrm{ph}}-\gamma_{\rm X}\right)^2}.    
\end{equation}
If we account for the detuning between the photonic end excitonic modes $\Delta \Omega_{\Gamma}$:
\begin{equation}
    \tilde{E}_{\rm UP/LP} = \frac{E_{\rm ph}+E_{\rm X}}{2}-i\frac{\gamma_{\rm ph}+\gamma_{\rm X}}{2}{}\pm\sqrt{g_2^2+\big(\frac{E_{\rm ph}-E_{\rm X}}{2}-i\frac{\gamma_{\rm ph}-\gamma_{\rm X}}{2}\big)^2}
\end{equation}
than the Rabi splitting is:
\begin{equation}
    \Omega_R = 2\rm{Re}\big(\sqrt{g_2^2+\Big(\frac{E_{\rm ph}-E_{\rm X}}{2}-i\frac{\gamma_{\rm ph}-\gamma_{\rm X}}{2}\big)^2}~\Big)
\end{equation}

\begin{table*}[h!]
\centering
\caption{
Exciton-photon coupling strength at the $\Gamma$ point for Type 1 and Type 2 structures.
The bare photonic-mode energy $\operatorname{Re}E_{\rm ph}$, detuning $\Delta\Omega=\operatorname{Re}E_{\rm ph}-E_X$, photonic and excitonic linewidths $\gamma_{\rm ph}$ and $\gamma_{\rm ex}$, Rabi splitting $\Omega_R$, and coupling energies $g_1$ and $g_2$ are presented in meV.
The coupling $g_1$ is estimated neglecting the detuning, while $g_2$ includes the detuning correction.
}
\label{tab:gamma_coupling_detuning}

\begin{tabular*}{0.8\textwidth}{@{\extracolsep{\fill}}|c|c|c|c|c|c|c|c|c|}
\hline
$h_{\rm hBN}$ 
& Type 
& $\operatorname{Re}E_{\rm ph}$ 
& $\Delta\Omega_{\Gamma}$ 
& $\gamma_{\rm ph}$ 
& $\gamma_{\rm ex}$ 
& $\Omega_R$ 
& $g_1$ 
& $g_2$ 
\\
\hline
\hline

\multirow{2}{*}{6}
& 1 & 1643.465 &  1.465 & 0.826   & 1.000 & 3.435 & 1.720 & 1.556 \\
& 2 & 1642.107 &  0.107 & 0.00006 & 1.000 & 2.897 & 1.532 & 1.531 \\

\hline

\multirow{2}{*}{90}
& 1 & 1640.731 & -1.269 & 2.201 & 1.000 & 7.779 & 3.936 & 3.884 \\
& 2 & 1643.167 &  1.167 & 0.005 & 1.000 & 7.824 & 3.944 & 3.900 \\

\hline

\multirow{2}{*}{120}
& 1 & 1640.609 & -1.391 & 2.297 & 1.000 & 8.668 & 4.382 & 4.327 \\
& 2 & 1642.744 &  0.744 & 0.010 & 1.000 & 8.616 & 4.336 & 4.320 \\

\hline
\end{tabular*}
\end{table*}

\end{document}